\documentclass[12pt, %a4 paper
]{article}
\usepackage{booktabs}

\usepackage[utf8]{inputenc}
\usepackage{amssymb}
\usepackage{amsmath}

\usepackage{geometry}
\newtheorem{theorem}{Theorem}%[chapter]

\newtheorem{corollary}{Corollary}%[chapter]
\newtheorem{definition}{Definition}%[chapter]
\newtheorem{lemma}{Lemma}%[chapter]
\newtheorem{remark}{Remark}%[chapter]
\newenvironment{proof}[1][Proof]{\emph{#1.} }{\  \hfill $\square $ \vspace{5 pt}}

\usepackage{natbib}
\usepackage{amssymb}
\usepackage[dvips]{graphics}
\usepackage{amsmath}

\usepackage{mathpazo}
\usepackage{enumerate}

\usepackage{amsfonts}

\usepackage{amssymb}

\usepackage{enumerate}

\usepackage{mathrsfs}

\usepackage[usenames]{color}
\usepackage{cancel}

\usepackage{pgf,tikz}

\usetikzlibrary{decorations.pathreplacing,angles,quotes}
\usetikzlibrary{arrows.meta}
\usetikzlibrary{arrows, decorations.markings}
\tikzset{myptr/.style={decoration={markings,mark=at position 1 with %
       {\arrow[scale=2,>=stealth]{>}}},postaction={decorate}}}
\usepackage{hyperref}
\hypersetup{
    colorlinks,
    citecolor=blue,
    filecolor=black,
    linkcolor=blue,
    urlcolor=blue
}
\usepackage{pgf,tikz}
\usetikzlibrary{arrows}

\usepackage{fancyhdr}
\usepackage{soul}

\usepackage{emptypage}
\usepackage[T1]{fontenc}
\DeclareFontFamily{T1}{calligra}{}
\DeclareFontShape{T1}{calligra}{m}{n}{<->s*[1.44]callig15}{}
\DeclareMathAlphabet\mathcalligra   {T1}{calligra} {m} {n}

\usepackage{multirow}
\usepackage{array}
\usepackage{mathtools}
\usepackage{arydshln}
\usepackage{threeparttable}
\usepackage{booktabs,caption}

\usepackage{ifthen}
\newboolean{showcomments}
\setboolean{showcomments}{true}

\newcommand{\pablo}[1]{  \ifthenelse{\boolean{showcomments}}
{\textcolor{green!50!black}{(T: #1)}}{}}
\newcommand{\marcelo}[1]{\ifthenelse{\boolean{showcomments}}
{\textcolor{red}{(M: #1)}}{}}
\newcommand{\agustin}[1]{  \ifthenelse{\boolean{showcomments}}
{\textcolor{blue!50!black}{(T: #1)}}{}}

\begin{document}

\title{Trade-off between manipulability and dictatorial power: a proof of  the Gibbard--Satterthwaite Theorem\thanks{%
%Thanks to be added. 
I thank Pablo Arribillaga, Jordi Massó, Pablo Neme, the Associate Editor, and two anonymous referees for their helpful comments. I acknowledge financial support
from UNSL through grants 032016, 030120, and 030320, from Consejo Nacional
de Investigaciones Cient\'{\i}ficas y T\'{e}cnicas (CONICET) through grant
PIP 112-200801-00655, and from Agencia Nacional de Promoción Cient\'ifica y Tecnológica through grant PICT 2017-2355.}}


\author{Agustín G. Bonifacio\thanks{%{\scriptsize 
Instituto de Matem\'{a}tica Aplicada San Luis (UNSL and CONICET) and Departamento de Matemática, Universidad Nacional de San Luis, San Luis, Argentina. E-mail: \href{mailto:agustinbonifacio@gmail.com}{\texttt{abonifacio@unsl.edu.ar}} 
} }

\date{\today}

\maketitle

\begin{abstract}
By endowing the class of \emph{tops-only} and \emph{efficient} social choice rules with a dual order structure that exploits the trade-off between  different degrees  of manipulability  and dictatorial power rules allow agents to have, we provide a proof of the Gibbard--Satterthwaite Theorem.

\bigskip

\bigskip


\end{abstract}

%\newpage
\section{Introduction}

The Gibbard--Satterthwaite Theorem states that, when more than two alternatives and all possible preferences over alternatives are considered, a social choice rule is unanimous and strategy-proof if and only if it is dictatorial  \citep{gibbard73, satterthwaite1975strategy}. There are several interesting proofs of the Gibbard--Satterthwaite Theorem \citep[see Section 3.3 in ][and references therein]{barbera2011strategyproof}. In this paper, we provide a new proof of this theorem.
To do this, we exploit the trade-off between the different degrees of  manipulability  and the different degrees of dictatorial power  that rules allow agents to have within the  class of tops-only  and efficient rules.  

The idea behind our proof is the following. First, we show that each unanimous and strategy-proof rule is both tops-only and efficient.  
Since dictatorial rules are tops-only and efficient as well, it is safe to restrain our analysis to this class of rules in order to prove Gibbard--Satterthwaite's result.

The novelty of our approach consists of defining two orders to compare rules. One order compares rules according to their manipulability. A profile of preferences is manipulable for a given tops-only rule if there is an agent who can manipulate the rule at a (possibly different) profile in which this agent has the same top and each remaining agent has the same preference as in the original profile. A rule is at least as manipulable as another rule if the former has as many manipulable profiles as the latter.\footnote{\cite{maus2007anonymous} compare rules according to a similar criterion. Other comparison criteria are studied, for example, in \cite{arribillaga2016comparing} and \cite{pathak2013school}.} Therefore, since strategy-proof rules have no manipulable profiles, any rule is at least as manipulable as a strategy-proof rule. The other order compares rules according to the dictatorial power of the agents.  A rule is at least as dictatorial as another rule if the former has as many dictatorial profiles as the latter, where a dictatorial profile is such that agents not obtaining their top alternative cannot unilaterally affect the social outcome by changing their preference. Therefore, since dictatorial rules have all of their profiles dictatorial, any dictatorial rule is at least as dictatorial as any other rule.

The crucial fact is that, given a rule,  each preference profile  
is either manipulable or dictatorial for that rule, which is equivalent to saying that both orders are dual. Gibbard--Satterthwaite's result follows easily from this.  Clearly, every dictatorial rule is strategy-proof. 
To see the converse, consider a strategy-proof 
rule. 
As we already pointed out, any other rule is at least as manipulable as this strategy-proof  
rule. Then, 
by the duality between the orders, this strategy-proof  
rule is at least as dictatorial as any other rule. Thus, this strategy-proof rule is dictatorial.

The paper is organized as follows. Preliminaries are introduced in Section \ref{preliminaries},  where it is also shown that every unanimous and strategy-proof rule is also tops-only and efficient. Then, in Section \ref{section dual}, the dual order structure of tops-only and efficient rules together with the proof of the Gibbard--Satterthwaite Theorem are presented.

\section{Preliminaries}\label{preliminaries}

A set of \textit{agents }$N=\{1,\ldots ,n\}$, with $|N|\geq 2$, has to choose
an alternative from a finite set $X$, with  $\left\vert
X\right\vert \geq 3$. Each agent $i\in N$ has a strict \textit{%
preference} $P_{i}$ over $X.$ Denote by $t(P_{i})$ to the
best alternative according to $P_{i}$, called the 
\textit{top }of $P_{i}$.
Given $i \in N$ and $x \in X$, a generic preference for $i$ with a top equal to $x$ is denoted by $P_i^x$. We denote by $R_{i}$ the weak preference over $X$
associated to $P_{i}$. % \textit{i.e., }for all $x,y\in X$, $xR_{i}y$ if and only if either $x=y$ or $xP_{i}y.$ 
Let $\mathcal{P}$ be the (universal) domain of all
strict preferences over $X.$ A \textit{(preference) profile} is an ordered list of $n$ preferences, %$n$-tuple 
 $P=(P_{1},\ldots ,P_{n})\in \mathcal{P}^{n},$  one for each agent. Given a profile $P$ and a set of agents $S$, $P_{-S}$ denotes the subprofile in $\mathcal{P}^{n-|S|}$ obtained by deleting  each $P_{i}$ for $i \in S$ from $P$.

A \textit{(social choice) rule} is a function $f:\mathcal{P}%
^{n}\longrightarrow X$ that selects an alternative in $X$ for each preference profile in $%
\mathcal{P}^{n}$. Let $\mathcal{F}$ denote the set of all rules. We assume throughout that   rules are \textit{unanimous}, i.e., for each $P \in \mathcal{P}^n$ such that $t(P_i)=x$ for each $i \in N,$  $f(P)=x$. 
Given a rule $f:\mathcal{P}^{n}\longrightarrow X$, a profile $P \in \mathcal{P}^n$ and a preference $P_i'\in \mathcal{P}$,  we say that \emph{agent $i$ manipulates $f$ at $P$ via $P_i'$} if $f(P_i',P_{-i})P_if(P)$. If no agent ever manipulates $f$, then $f$ is \emph{strategy-proof}. The set of all \emph{strategy-proof} and \emph{unanimous} rules is denoted by $\mathcal{S}$. A rule $f:\mathcal{P}%
^{n}\longrightarrow X$ is \textit{dictatorial} if there is $i\in N$ (the \emph{dictator}) such
that, for each $P\in \mathcal{P}^{n}$, $f(P)=t(P_i)$.
 The set of all \emph{dictatorial} rules is denoted by $\mathcal{D}$. The Gibbard--Satterthwaite Theorem, then, states that $\mathcal{S}=\mathcal{D}$.

A  rule $f:\mathcal{P}^{n}\longrightarrow X$ is \textit{efficient} if, for each $P\in \mathcal{P}^{n}$,
there is no $x\in X$ such that $xP_{i}f(P)$ for each $i\in N$. Next, we prove that \emph{unanimous} and \emph{strategy-proof} rules are \emph{efficient} as well.  

\begin{lemma}\label{lemma efficient}
If $f \in \mathcal{S}$, then $f$ is  efficient.
%$\mathcal{S} \subseteq \mathcal{T}.$
\end{lemma}
\begin{proof}
    Let $f \in \mathcal{S}$. Assume that it is not \emph{efficient}. Then, there are $x \in X$ and $P \in \mathcal{P}^n$ such that $xP_if(P)$ for each $i \in N$. Let $P^\star \in \mathcal{P}^n$ be such that, for each $i \in N$, $P^\star_i:x,f(P),\ldots$ 
    By \emph{strategy-proofness}, $f(P_1^\star, P_{-1})\neq x$ (otherwise  agent 1 manipulates $f$ at $P$ via $P_1^\star$). Furthermore, again by \emph{strategy-proofness},  $f(P_i^\star, P_{-i})=f(P)$ (otherwise, since $f(P_i^\star, P_{-i})\neq x$,  agent 1 manipulates $f$ at $(P_i^\star, P_{-i})$ via $P_i$). Using the same argument, changing the preference of one agent at a time, it follows that $f(P^\star)=f(P)\neq x$. This contradicts \emph{unanimity}. Thus, $f$ is efficient. 
    \end{proof}

A rule $f:\mathcal{P}^{n}\longrightarrow X$ is 
\textit{tops-only }if for each $P,P^{\prime }\in \mathcal{P}^{n}$
such that $t(P_{i})=t (P_{i}')$ for each $i\in N$, 
$f(P)=f(P')$. Let $\mathcal{T}$ denote the set of all \emph{tops-only} rules. It turns out that \emph{unanimous} and \emph{strategy-proof} rules are \emph{tops-only} as well.

  \begin{lemma}\label{lemma SP implies tops-only}
If $f \in \mathcal{S}$, then $f \in \mathcal{T}.$
%$\mathcal{S} \subseteq \mathcal{T}.$
\end{lemma}
\begin{proof}
Let $f \in \mathcal{S}$. Since domain $\mathcal{P}$ is minimally rich and satisfies Property $T^\star$, then $f \in \mathcal{T}$ by Theorem 2 in \cite{chatterji2011tops}.\footnote{A domain (of preferences)  $\mathcal{D} \subseteq \mathcal{P}$ is \textbf{minimally rich} if for each $x \in X$ there is $P_i \in \mathcal{D}$ such that $t(P_i)=x.$ A domain  $\mathcal{D} \subseteq \mathcal{P}$ satisfies \textbf{Property $\boldsymbol{T^\star}$} if for each $P_i \in \mathcal{D}$ and each $x \in X \setminus \{t(P_i)\}$ it holds that, for each $\overline{y} \in \{y \in X :  yP_i'x$ for each $P_i'$ such that $t(P_i')=t(P_i)\}$ there is $\overline{P}_i \in \mathcal{D}$ such that: (i) $t(\overline{P}_i)=x$ and (ii) $\overline{y}\overline{P}_i \overline{z}$ for each $\overline{z} \in \{z \in X: xP_iz\}$ \citep[see Definition 9 in][for more detail]{chatterji2011tops}. It is straightforward to check that the universal domain $\mathcal{P}$ satisfies both properties.}   
\end{proof}

Let $\mathcal{TE}$ denote the set of all \emph{tops-only} and \emph{efficient} rules. It is clear that \emph{dictatorial} rules are \emph{efficient} and \emph{tops-only}, i.e., $\mathcal{D} \subseteq \mathcal{TE}$. Moreover, by Lemmata \ref{lemma efficient} and \ref{lemma SP implies tops-only} the next result follows.

\begin{corollary}\label{lemma tops-only}
$\mathcal{S} \subseteq \mathcal{TE}.$
\end{corollary}

\section{Dual order structure and the proof}\label{section dual}

Before presenting the comparability criteria, we state a result that says that rules in $\mathcal{TE}$ always select one of the top alternatives in each profile of preferences. Given $x \in X$ and $P \in \mathcal{P}^n$, let $N_x(P) \equiv \{i \in N : t(P_i)=x\}$.

%Before presenting the comparability criteria, we state a result analogous to Lemma \ref{Lema sale un top} but now for rules  in $\mathcal{T}$.% always  select a top alternative in each profile. 

\begin{lemma}\label{lemma eff implica algun top}
    %If $f \in \mathcal{T}$, then for each $P \in \mathcal{P}^n$, $f(P)=t(P_i)$ for some $i \in N$. 
    Let $f \in \mathcal{TE}$, $P \in \mathcal{P}^n$ and $x \in X$ be such that $f(P)=x$. Then, $N_x(P)\neq \emptyset$.
\end{lemma}
\begin{proof} Let $f \in \mathcal{TE}$ and assume there is $P \in \mathcal{P}^n$ such that $t(P_i)\neq f(P)$ for each $i \in N$. Let $P^\star \in \mathcal{P}^n$ be such that, for each $i \in N$, $t(P_i^\star)=t(P_i)$ and $y P_i^\star f(P)$ for each $y  \in X \setminus \{f(P)\}$. By \emph{tops-only}, $f(P^\star)=f(P)$. Let $x \in X \setminus \{f(P)\}$. Then, $xP_i^\star f(P^\star)$ for each $i \in N$, contradicting \emph{efficiency}. Therefore, $N_x(P)\neq \emptyset$. 
\end{proof}

%\textcolor{blue}{A profile of preferences is manipulable for a given tops-only rule if there are an agent and a (possibly different) profile in which this agent has a preference with the same top and each remaining agent has the same preference as in the original profile, such that this agent manipulates the rule at this new profile. }

A profile of preferences is manipulable for a given \emph{tops-only} rule if there is an agent who can manipulate the rule at a (possibly different) profile in which this agent has the same top and each remaining agent has the same preference as in the original profile. Formally, let $f \in \mathcal{T}$ and let $P\in \mathcal{P}^n$. Profile $P$ is \emph{manipulable for $f$} if there are $i \in N$ and  $P_i^\star, P_i' \in \mathcal{P}$ such that $t(P_i^\star)=t(P_i)$ and $i$ manipulates $f$ at $(P_i^\star,P_{-i})$ via $P_i'$.\footnote{Note that for this definition it is crucial that $f$ be \emph{tops-only}.} Denote by $M_f$  the set of all manipulable profiles for $f$. Our comparability criterion with respect to manipulability is presented next. 

\begin{definition}
Let $f,g \in  \mathcal{T}.$  We say that \textbf{$\boldsymbol{f}$ is at least as manipulable as $\boldsymbol{g}$,} and write $\boldsymbol{f\succeq_m g}$,  if $|M_f|\geq|M_g|$.
\end{definition}

%The relation $\succeq_m$ is complete (and therefore reflexive) and transitive.

%Remember that, by Corollary \ref{lemma tops-only}, every \emph{unanimous} and \emph{strategy-proof} rule  belongs to $\mathcal{T}$. 

It is clear that  $f \in \mathcal{T}$ is \emph{strategy-proof} if and only if  $M_f=\emptyset$. Thus, the next remark follows.

\begin{remark}\label{remark strategy-proof}\rm
Let $f \in  \mathcal{T}.$ Then, $f$ is strategy-proof if and only if  $g \succeq_m f$ for each $g \in  \mathcal{T}.$  
\end{remark}

A profile of preferences is dictatorial for a given (not necessarily \emph{tops-only}) rule if agents not obtaining their top alternative cannot unilaterally affect the social outcome by changing their preference. Formally, let $f \in  \mathcal{F}$ and let $P\in \mathcal{P}^n$. Profile $P$ is \emph{dictatorial for $f$} if, for each $i \in N$ such that $t(P_i)\neq f(P)$, we have  $f(P_i', P_{-i})=f(P)$ for each $P_i' \in \mathcal{P}$.\footnote{Note that a unanimous profile is trivially dictatorial for any rule.} Denote by $D_f$  the set of all dictatorial profiles for $f$. Our comparability criterion with respect to dictatorial power is presented next.  

\begin{definition}
Let $f,g \in \mathcal{F}$. We say that \textbf{$\boldsymbol{f$ is at least as  dictatorial as $g}$,} and write $\boldsymbol{f\succeq_d g}$,  if $|D_f|\geq|D_g|$. 
\end{definition}

%The relation $\succeq_d$ is complete (and therefore reflexive) and transitive.

The only rules in $\mathcal{TE}$ for which all profiles are dictatorial are precisely \emph{dictatorial} rules. 

\begin{lemma}\label{lema dictatorial}
  Let $f \in \mathcal{TE}$. Then, $f \in \mathcal{D}$ if and only if  $D_f=\mathcal{P}^n$.   
\end{lemma}
\begin{proof}
    It is clear that if $f$ is \emph{dictatorial}, then $D_f=\mathcal{P}^n$. Let $f \in \mathcal{TE}$ be such that $D_f=\mathcal{P}^n$. We start with the case $n=2$. If $f$ is not \emph{dictatorial} then for each $i \in \{1,2\}$ there is $P^i \in \mathcal{P}^2$ such that $f(P^i)\neq t(P^i_i)$. Thus, by Lemma \ref{lemma eff implica algun top},  
        \begin{equation}\label{eq lema dictatorial}
         f(P^1_1,P^1_2)=t(P^1_2) \text{ and } f(P^2_1,P^2_2)=t(P^2_1).  
        \end{equation}
    %Consider now profile $(P^1_1,P^2_2)$. 
    There are two cases to consider:
    
    \noindent $\boldsymbol{1}$. $\boldsymbol{t(P^1_1)\neq t(P^2_2)}$. By Lemma \ref{lemma eff implica algun top}, w.l.o.g., we have  $f(P^1_1, P^2_2)=t(P^1_1)$. Since profile $(P^1_1, P^2_2)$ is dictatorial and $f(P^1_1, P^2_2)\neq t(P^2_2)$, it follows that $f(P^1_1, P^1_2)=t(P^1_1)$, contradicting \eqref{eq lema dictatorial}. 
    
    \noindent $\boldsymbol{2}$. $\boldsymbol{t(P^1_1)= t(P^2_2)}$. Let $\widetilde{P}_1 \in \mathcal{P}$ be such that $t(\widetilde{P}_1) \notin \{t(P^1_1), t(P^1_2)\}$. Since profile $(P^1_1, P^1_2)$ is dictatorial and $f(P^1_1, P^1_2)\neq t(P^1_1)$, it follows that $$f(\widetilde{P}_1, P^1_2)=t(P^1_2) \text{ and } f(P^2_1, P^2_2)=t(P^2_1).$$ Then, as $t(\widetilde{P}_1)\neq t(P^2_2)(=t(P^1_1))$, the argument follows as in the previous case, and we reach another contradiction. Hence, $f$ is dictatorial when $n=2$. 
    
    Assume now that $n \geq 3$ and that every $n-1$ agent rule $f' \in \mathcal{TE}$ such that $D_{f'}=\mathcal{P}^{n-1}$ is \emph{dictatorial}. Given $f:\mathcal{P}^n \longrightarrow X$ define rule $g:\mathcal{P}^{n-1} \longrightarrow X$ as follows.\footnote{The idea of defining a $n-1$ agent rule from an $n$ agent rule by ``coalescing'' two agents is due to \cite{sen2001another}.} For each $(P_1,P_3,\ldots,P_n) \in \mathcal{P}^{n-1},$ $$g(P_1,P_3,\ldots,P_n)=f(P_1,P_1,P_3, \ldots,P_n).$$ Clearly, $g$ is \emph{efficient}, $\emph{tops-only}$, and $D_g=\mathcal{P}^{n-1}$. Therefore, $g$ has a dictator, say $i^\star$.  First, assume $i^\star \in \{3,\ldots, n\}$. We now show that $i^\star$ is also the dictator of $f$. If $i^\star$ is not the dictator of $f$, then by Lemma \ref{lemma eff implica algun top} there is $(P_1,P_2,P_3, \ldots, P_n) \in \mathcal{P}^n$ and $k \in N \setminus \{i^\star\}$ such that $f(P_1,P_2,P_3, \ldots, P_n)=t(P_k)\neq t(P_{i^\star})$. Since $f$ is \emph{tops-only} and profile $P$ is   dictatorial,   $f(P_k,P_2,P_3, \ldots,P_n)=t(P_k)$. Again,  since $f$ is \emph{tops-only} and $P$ is   dictatorial,   $f(P_k,P_k,P_3, \ldots,P_n)=t(P_k)$. By the definition of $g$, $g(P_k, P_3, \ldots,P_n)=t(P_k)$, contradicting that $i^\star$ is the dictator of $g$. Therefore, $i^\star$ is the dictator of $f$ as well. Next, assume $i^\star=1$. Notice that  $f(P_1,P_2,P_3, \ldots, P_n) \in \{t(P_1), t(P_2)\}$ for each $(P_1,P_2,P_3, \ldots, P_n) \in \mathcal{P}^n$ (otherwise, there is $j \in \{3,\ldots,n\}$ such that $f(P_1,P_2,P_3, \ldots, P_n)=t(P_j) \notin \{t(P_1),t(P_2)\}$ and, since $(P_1,P_2,P_3, \ldots, P_n)$ is dictatorial, $f(P_1,P_1,P_3, \ldots, P_n)=t(P_j)$, contradicting that $1$ is the dictator of $g$). Let $(\overline{P}_3, \ldots, \overline{P}_n) \in \mathcal{P}^{n-2}$. Define rule $h:\mathcal{P}^2 \longrightarrow X$ as follows. For each $(P_1,P_2) \in \mathcal{P}^2$, $$h(P_1,P_2)=f(P_1,P_2,\overline{P}_3, \ldots, \overline{P}_n).$$ Clearly, $h$ is \emph{efficient}, $\emph{tops-only}$, and $D_h=\mathcal{P}^{2}$. Therefore, as we already proved for the case $n=2$, $h$ has a dictator. It remains to be shown that this dictator does not depend on the sub-profile $(\overline{P}_3, \ldots, \overline{P}_n)$ chosen to define $h$. Assume it does. Then, there is another sub-profile $(\widetilde{P}_3, \ldots, \widetilde{P}_n) \in \mathcal{P}^{n-1}$ such that, w.l.o.g.,
    \begin{equation}\label{eq lema dictatorial 2}
      f(P_1,P_2,\overline{P}_3, \ldots, \overline{P}_n)=t(P_1) \text{ and } f(P_1,P_2,\widetilde{P}_3, \ldots, \widetilde{P}_n)=t(P_2), 
    \end{equation}
    and also $t(P_1)\neq t(P_2)$. Let $z \in X\setminus \{t(P_1),t(P_2)\}$ and consider any $P^z_3 \in \mathcal{P}$. Then, $$f(P_1,P_2,P_3^z,\overline{P}_{-1,2,3}) \in \{t(P_1),z\}$$  (otherwise, since $(P_1,P_2,P_3^z,\overline{P}_{-1,2,3})$ is dictatorial, $f(P_1,P_2,P_3^z,\overline{P}_{-1,2,3}) \notin \{t(P_1),z\}$ implies $f(P_1,P_2,\overline{P}_3, \ldots, \overline{P}_n) \notin \{t(P_1),z\}$, a contradiction). Using the same argument, changing the preference of one agent at a time, it follows that $f(P_1,P_2,P_{-1,2}^z) \in \{t(P_1),z\}$.  Similarly, but now starting from $f(P_1,P_2,\widetilde{P}_3, \ldots, \widetilde{P}_n)$, we get $f(P_1,P_2,P_{-1,2}^z) \in \{t(P_2),z\}$.  Thus, $f(P_1,P_2,P_{-1,2}^z)=z$ and, since profile $(P_1,P_2,P_{-1,2}^z)$ is dictatorial, $f(P_1,P_1,P_{-1,2}^z)=z$. Then, $g(P_1,P_{-1,2}^z)=z$ and agent $1$ is not the dictator of $g$, a contradiction. Therefore, either agent 1 or agent 2 is the dictator of $f$. 
\end{proof}

Note that, within $\mathcal{T}$, \emph{efficiency} is necessary for Lemma \ref{lema dictatorial} to hold. Otherwise, a constant function $f$ also satisfies that all profiles are dictatorial for $f$. The next remark follows from Lemma \ref{lema dictatorial}.

\begin{remark}\label{remark dictatorial} \rm
Let $f \in \mathcal{TE}$. Then,  $f \in \mathcal{D}$ if and only if  $f \succeq_d g$ for each $g \in \mathcal{TE}$. 
\end{remark}

It turns out that the classification of a preference profile as manipulable or dictatorial, for a given rule in $\mathcal{T}$, is exhaustive.

\begin{lemma}\label{lema dualidad}
    Let $f \in \mathcal{T}$ and let $P \in \mathcal{P}^n$. Then, profile $P$  is either dictatorial for $f$ or manipulable for $f$. 
\end{lemma}
\begin{proof}
    Let $f \in \mathcal{T}$ and let $P \in \mathcal{P}^n$.  Assume profile $P$ is not dictatorial for $f$. Then, there are $i \in N$ with  $t(P_i)\neq f(P)$ and $P_i' \in \mathcal{P}$ such that $f(P_i',P_{-i})\neq f(P)$. Consider $P^\star_i \in \mathcal{P}$ such that $t(P_i^\star)=t(P_i)$ and $f(P_i',P_{-i})P_i^\star f(P)$. By \emph{tops-only}, $f(P_i^\star,P_{-i})=f(P)$. Thus, $f(P_i',P_{-i})P_i^\star f(P_i^\star,P_{-i})$ and $i$ manipulates $f$ at $(P_i^\star,P_{-i})$ via $P_i'$. Therefore, $P$ is manipulable for $f$.  
\end{proof}

%The following example shows that \emph{efficiency} is necessary for the previous result.  

%\begin{example} \rm
 %   Let $N=\{1,2\}$ and $X=\{x,y,z\}$ with $x<y<z$. Let $f^y:\mathcal{P}^2 \longrightarrow X$ be defined by $f^y(P_1,P_2)=\textrm{med}_{<}\{t(P_1),t(P_2),y\}$. Consider profile $\widehat{P} \in \mathcal{P}^2$ such that $x \widehat{P}_1 y \widehat{P}_1 z$ and $z \widehat{P}_2 x \widehat{P}_2 y$. Then, $f^y(\widehat{P})=y$ so $f$ is not \emph{efficient}. Furthermore, if $\widetilde{P}_2\in \mathcal{P}$ is such that $t(\widetilde{P}_2)=y$, then $f^y(\widehat{P}_1,\widetilde{P}_2)=y$ and profile $(\widehat{P}_1,\widetilde{P}_2)$ is neither manipulable nor dictatorial for $f^y$.   \hfill $\Diamond$
%\end{example}

The dual order structure of $\mathcal{T}$ is an immediate consequence of Lemma \ref{lema dualidad}.

\begin{corollary}\label{lemma dual}
    Let $f,g \in \mathcal{T}$. Then, $f\succeq_d g$ if and only if $g \succeq_m f$.  
\end{corollary}

%\subsection{The Gibbard--Satterthwaite theorem}

We are now finally  in a position to prove the Gibbard--Satterthwaite Theorem.

\begin{theorem}%\citep{gibbard73,satterthwaite1975strategy} 
$\mathcal{S}=\mathcal{D}.$     
\end{theorem}
\begin{proof}
    It is clear that $\mathcal{D} \subseteq \mathcal{S}$. Next, we prove that $\mathcal{S} \subseteq \mathcal{D}$. Let $f \in \mathcal{S}$. By Corollary \ref{lemma tops-only}, $f \in \mathcal{TE}$. By Remark \ref{remark strategy-proof}, $g \succeq_m f$ for each $g \in \mathcal{TE}$. By Corollary \ref{lemma dual}, $f\succeq_d g$ for each $g \in \mathcal{TE}$. Thus, by Remark \ref{remark dictatorial}, $f \in \mathcal{D}$.   
\end{proof}

\bibliographystyle{ecta}
\bibliography{biblioGS}

\end{document}